\documentclass[fleqn,usenatbib]{mnras}
\usepackage{amsmath}	
\usepackage{newtxtext,newtxmath}
\usepackage[T1]{fontenc}
\usepackage{ae,aecompl}

\usepackage{graphicx}	
\usepackage{amssymb}	
\usepackage{bm}
\usepackage{listings}
\usepackage{algorithm2e}

\title[Planet-disc interactions with Discontinuous Galerkin Methods using GPUs]{Planet-disc interactions with Discontinuous Galerkin Methods using GPUs}

\author[David A. Velasco Romero et al.]{%
David A. Velasco Romero$^{1,2,3}$\thanks{E-mail: david.velasco@icf.unam.mx}, Maria Han Veiga$^{2,4}$, Romain Teyssier$^{2}$ and
\newauthor Fr\'ed\'eric S. Masset$^{3}$
\\
$^{1}$Universidad Aut\'onoma del Estado de Morelos, Av. Universidad s/n, 62210 Cuernavaca, Mor., Mexico\\
$^{2}$Institute of Computational Science, University of Zurich\\
$^{3}$Instituto de Ciencias F\'isicas, Universidad Nacional Aut\'onoma de M\'exico, Av. Universidad s/n, 62210 Cuernavaca, Mor., Mexico\\
$^{4}$Institute of Mathematics,
University of Zurich\\
}

\date{Accepted 2018 May 2. Received 2018 April 27; in original form 2018 March 6}

\pubyear{2018}

\begin{document}
\label{firstpage}
\pagerange{\pageref{firstpage}--\pageref{lastpage}}
\maketitle

\begin{abstract}
We present a two-dimensional Cartesian code based on high order discontinuous Galerkin methods, implemented to run in parallel over multiple GPUs. A simple planet-disc setup is used to compare the behaviour of our code against the behaviour found using the FARGO3D code with a polar mesh. We make use of the time dependence of the torque exerted by the disc on the planet as a mean to quantify the numerical viscosity of the code. 
We find that the numerical viscosity of the Keplerian flow can be as low as a few $10^{-8}r^2\Omega$, $r$ and $\Omega$ being respectively the local orbital radius and frequency, for fifth order schemes and resolution of $\sim 10^{-2}r$. Although for a single disc problem a solution of low numerical viscosity can be obtained at lower computational cost with FARGO3D (which is nearly an order of magnitude faster than a fifth order method), discontinuous Galerkin methods appear promising to obtain solutions of low numerical viscosity in more complex situations where the flow cannot be captured on a polar or spherical mesh concentric with the disc.
\end{abstract}




\section{Introduction}
The discovery of exoplanetary systems at an ever increasing pace has triggered a lot of theoretical works to understand and account for their extraordinary diversity. A significant fraction of these studies has been undertaken through intensive computational simulations in which protoplanets grow and gravitationally interact with their parent disc. The more common practice for simulations of planet-disc interactions is through grid-based codes. Among the plethora of such codes used for studies of planet-disc interactions we can cite Athena \citep[e.g.][]{2014ApJ...785..122Z}, DISCO \citep{2016ApJS..226....2D}, FARGO and FARGO3D \citep{2000A&AS..141..165M,2016ApJS..223...11B}, NIRVANA \citep{2003ApJ...586..540D}, PENCIL \citep{2009A&A...493.1125L}, PEnGUIn \citep{2014ApJ...782...88F}, PLUTO \citep{2012A&A...545A.152M} and RODEO \citep{2006A&A...450.1203P}. Some of these codes are relatively new, while others have been used for over a decade. The properties and performance of the latter have been studied by \citet{2006MNRAS.370..529D} in a code comparison project dedicated to planet-disc interactions. By far the most common meshes are polar meshes centered on the primary (in two-dimensions) or cylindrical or spherical meshes (also centered on the primary, and coplanar with the disc) for three-dimensional simulations. There are very few exceptions to this, such as the studies of \citet{2008MNRAS.386..164P}, who performed short-term simulations of the fast migration of giant planets. Cylindrical or spherical meshes are naturally adapted to the geometry of the problem at hand, and result in much smaller numerical viscosity of the disc's flow than their Cartesian counterpart, for a given scheme and cell size. On the other hand planet-disc interactions are very sensitive to the disc's viscosity, be it through the saturation of the corotation torque in the low-mass regime \citep{2001ApJ...558..453M,2010ApJ...723.1393M,2011MNRAS.410..293P} or through the gap opening processes for giant planets \citep{1986ApJ...307..395L,2006Icar..181..587C,2014ApJ...782...88F}.  
There is a growing body of evidence that protoplanetary discs have a low effective viscosity, if any at all. The inclusion of non-ideal MHD effects in theoretical models of protoplanetary discs lead to a qualitatively different picture from earlier models, and suggest that the flow is laminar over most of the disc \citep{2013ApJ...769...76B,2014A&A...566A..56L}, while attempts of detection of turbulent motion in nearby protoplanetary discs lead to ever decreasing upper limits \citep[e.g.][]{2018ApJ...856..117F}. Studies of planet-disc interactions should therefore be undertaken with schemes of very low numerical viscosity. Not all numerical studies of protoplanetary discs or environments can be done on cylindrical or spherical meshes, however. As they grow in complexity and realism, they may be better done on Cartesian meshes with AMR \citep{2015A&A...579A..32L,2017A&A...599A..86H}. This can also happen if several discs are considered at the same time, such as the circumstellar discs of a multiple star. Under such circumstances, reaching the very low levels of viscosity required to capture correctly the interaction between the disc and its forming planets may prove challenging. Recently, \citet{2015MNRAS.453.4278S} presented an implementation of discontinuous Galerkin (DG) schemes aimed at describing astrophysical flows. The high order of the solutions provided by these schemes suggests that they may be able to capture differentially rotating discs with a very low viscosity. Besides, these schemes present the interesting property that they conserve angular momentum to machine accuracy in the parts of the flow where no limiting occurs. This property is highly desirable for long-term simulations of planet-disc interactions, where most of the planet's drift can be accounted for by an exchange of angular momentum between the planet and its coorbital region: a spurious change of the angular momentum of the latter may induce an erroneous migration rate of the former. Since discontinuous Galerkin methods are compute intensive and have a small stencil, they are well suited to massive multi-threaded platforms such as \emph{Graphics Processing Units} (GPUs). 

For all the reasons explained above, we have implemented a two-dimensional, Cartesian version of DG schemes on GPUs, and evaluated their properties on Keplerian flows with embedded, intermediate mass planets. Although an implementation of DG schemes in cylindrical or spherical coordinates would be feasible, we regard our present implementation as a proof of concept in the least favourable case. As we shall see, we are able to obtain very small numerical viscosities even in the case of a Cartesian mesh, which suggests that better results would be attainable for coordinate systems fitted to the geometry of the flow. Our paper is organized as follows: in section~\ref{sec:finite-el}, we recall the main features of DG schemes, and provide some details about our implementation in section~\ref{sec:implement}. We then check our code's behaviour and convergence properties on standard tests in section~\ref{sec:benchmarks} and we present our results for the problem of a planet embedded in a protoplanetary disc in section~\ref{sec:ppdisc}. We use the time behaviour of the corotation torque as a diagnostic to evaluate the effective viscosity of the disc. At finite viscosity, this torque tends toward a finite, constant value which depends on the effective viscosity, whereas it oscillates and tends to zero in inviscid discs. The  asymptotic torque therefore constitutes an accurate measure of the disc's effective viscosity, albeit somehow indirect. Note that although our method can accurately determine the effective numerical viscosity of a given scheme, the exact value may differ if another method is used. We finally draw our conclusions in section~\ref{sec:conclusions}.

\section{Principles of discontinuous Galerkin schemes}
\label{sec:finite-el}
\subsection{Governing equations}
The Euler equations describe how the velocity, pressure and density of a moving fluid are related under the influence of a source term. They form a n-dimensional system of hyperbolic partial differential equations that can be written as

\begin{equation}
\label{eq:euler}
\frac{\partial \bm{u}}{\partial t} + \sum_{i=1}^n \frac{\partial }{\partial x_i}\bm{f}_i(u) = S(\bm{u},\bm{x})
\end{equation}
where
\begin{equation*}
    \bm{u} = \begin{pmatrix}
     \rho \\  \rho v_x \\ \rho v_y \\ E 
    \end{pmatrix}, \quad
    \bm{f} =
    \begin{pmatrix}
     \rho v_x & \rho v_y  \\  \rho v_x^2 + p & \rho v_x v_y  \\ \rho v_x v_y & \rho v_y^2 + p \\ (E+p)v_x & (E+p)v_y
    \end{pmatrix}, \quad
    \bm{S} =
    \begin{pmatrix}
     0 \\ -\rho \frac{\partial}{\partial x} \Phi \\ -\rho \frac{\partial}{\partial y} \Phi  \\ -\rho \bm{v}\cdot \nabla \Phi
    \end{pmatrix}
\end{equation*}
for a 2-dimensional flow under the influence of a gravitational source term with  potential $\Phi$. Here $\bm{f}$ represents the matrix of fluxes $(\bm{f}_x,\bm{f}_y)$ and $\bm{S}$ the source term.

The unknown quantities are density $\rho$, velocity $\bm{v}=(v_x,v_y)$, pressure $p$, and total energy $E$. The total energy can be expressed in terms of the density of internal energy $e$ and kinetic energy of the fluid, $E = e+\frac{1}{2}\rho \bm{v}\cdot \bm{v}$. For an ideal gas, the system is closed with the equation of state
\begin{equation}
    p =  e (\gamma - 1),
\end{equation}
where $\gamma$ denotes the adiabatic index.

\subsection{Discontinuous Galerkin method}
We follow the method formulated by \citet{cs5}, which we summarize below for a 2-dimensional scalar conservation law defined on a cartesian grid.

Consider a regular 2-dimensional domain $\Omega \in \mathbb{R^2}$, approximated by non-overlapping rectangular elements 

\[K_{i,j} =  [x_{i-1/2},x_{i+1/2}] \times  [y_{j-1/2},y_{j+1/2}],\]

where $(i,j)$ indexes the rectangles. Furthermore, consider the local space $V(K)$ given by the set of 2-dimensional polynomials with degree of at most $N_p$ in $x$ and $y$. We denote $\{\phi_i\}_{i=0}^{N_p}$ to be the set of polynomial basis of the local space $V(K)$.

For every rectangle $K_{i,j}$, the local solution is expressed as\footnote{ We drop the rectangle indices $i,j$ when it's not important to specify them.}:

\[u^K_h(\bm{x},t) = \sum_{i=0}^{N_p} \hat{u}^K_i(t) \phi_i(\bm{x}).\]

In this work we use a \textit{modal representation} of the solution. This means that the numerical solution in element $K$ is represented by the linear coefficients of the basis functions $\hat{u}^K_i(t)$ for $i = 0, ..., N_p$. In particular, Legendre polynomials are chosen as the polynomial basis because they are orthogonal to each other, i.e: $\int \phi_i(\bm{x}) \phi_j(\bm{x}) d\bm{x} = \delta_{ij}$.

A modal coefficient $\hat{u}_i^K(t)$ is obtained with the $L^2$ projection of the solution $u(x)$ restricted to element $K$ on the orthogonal basis vector $\phi_i(\bm{x})$:
\[ \hat{u}_i^K(t) = \int_K u(\bm{x},t)\phi_i(\bm{x})d\bm{x} \]

The pointwise values of the solution (nodal values)  $u_h^K(\bm{x},t)$ can be recovered by:
\[ u_h^K(\bm{x},t) = \sum_{i=0}^{N_p} \hat{u}_i^K(t) \phi_i(\bm{x}). \]

Finally, the scalar global solution $u(\bm{x},t)$ is given by \emph{stitching} together the local solutions defined in each local subspace $V(K)$, which is formally expressed as a direct sum (denoted with $\bigoplus$):

\[u(\bm{x},t)\approx u_h(\bm{x},t) = \bigoplus_{K\in \Omega_h}  u_h^K(\bm{x},t).\]

The extension to a system of equations is done by repeating the treatment described above for each variable in the vector solution.

\subsubsection{Space discretization}
Using the notation above, we discretize Eq.~\eqref{eq:euler} in space using the discontinuous Galerkin method. For each time $t$, the approximate solution $u_h(\bm{x},t)$ is sought in the finite element space of discontinuous functions. The weak formulation of Eq.~\eqref{eq:euler} is attained by multiplying the equation by a smooth test function $v(\bm{x})$, integrating over a control volume $K$ and applying the divergence theorem: 

\begin{align*}
\frac{d}{dt} \int_{K} u(\bm{x},t) v(\bm{x}) d\bm{x} &+ \sum_{e\in \partial K}\int_e f(u(\bm{x},t))\cdot n_{e,K} v(x) d\Gamma \\ &- \int_K f(u(\bm{x},t))\cdot \nabla v(\bm{x}) d\bm{x} = \int_K S(u) v(\bm{x}) d\bm{x}.
\end{align*}
Here $n_{e,K}$ denotes the outward unit normal to the edge $e$.

The exact solution is replaced with the approximate solution $u_h(\bm{x})$, the test function $v(\bm{x})$ by $v_h(\bm{x})$ and the integrals from the weak formulation are replaced by a suitable quadrature, yielding the semi-discrete formulation of the discontinuous Galerkin method, written as:

\begin{align}
 u_h(t=0)&=P_{V_h}(u_0) \nonumber \\
 \frac{\mbox{d}}{\mbox{dt}}\int_K u_h(\bm{x},t)v_h(\bm{x})\mbox{d\bf{x}} = &-\sum_{e\in\partial K}\sum_{i=0}^L h_{e,K}(\bm{x}_i,t)v_h(\bm{x}_i) w_i |e| \nonumber \\ &+ \sum_{j=0}^M f(u(\bm{x}_j,t))\cdot \nabla v_h(\bm{x}_j) w_j |K|  \nonumber \\
 &+\sum_{j=0}^M S(u(\bm{x}_j,t)) v_h(\bm{x}_j) w_j |K| \nonumber \\ &\quad \forall v_h(\bm{x})\in V(K) \forall K \in \Omega_h, \label{eq:dg}
\end{align}

where $P_{V_h}(\cdot)$ denotes the $L^2$ projection of the initial data $u_0(\bm{x})$ into the space of finite elements $V_h$, $\{(\bm{x}_i,w_i)\}_{i=0}^{L,M}$ are sets of Gauss-Legendre quadrature points (with their respective weights $w$) with different number of points $L$ and $M$, for the edge and volume integrals, $|e|$ the length the edge and $|K|$ the area of the control volume. Furthermore, $f(u(\bm{x},t))\cdot n_{e,K}$ is replaced by $h_{e,K}(x_i,t)$, the numerical flux, which determines a unique solution at the interface shared between neighbouring elements.

\subsubsection{Time discretization}
Because we choose our local solution space to be a set of orthogonal polynomials, we have an expression for the evolution of each mode $\hat{u}^K_i$ independently of the other modes. The semi-discrete form \eqref{eq:dg} reduces the partial differential equation to an ordinary differential equation of the form:

\[ \frac{d}{dt} u_h = \mathcal{L}(u), \]

where $\mathcal{L}$ denotes the right hand side of \eqref{eq:dg}. A Strong Stability Preserving (SSP) Runge-Kutta (RK) time discretization is used \citep{RK}. The time marching algorithm is detailed in Algorithm \ref{algo:tvd}.

\begin{algorithm}
\label{algo:tvd}
\SetAlgoLined
\KwData{$w^0_h = P_{V_h}(w_0)$}
\KwResult{$w^{n+1}_h$}
\For{$n = 0, ... N-1$}{
    $w^{(0)}_h = w^n_h$;\\
    $h = \Delta t$;\\
    \For{ i = 0, ... k}{
    	$k_i = \mathcal{L}(t^n+c_i\cdot h,y_i+h(a_{i,1}k_1 + ... + a_{i,i-1}k_{i-1}))$;\\
        $w^{(i+1)}_h = w^{(i)}_h + h \sum_{j=1}^i b_j k_j $;
        }
   $w^{n+1}_h = w^{(k+1)}_h$;
}
 \caption{TVD RK time marching algorithm}
 \label{alg:english}
\end{algorithm}

The coefficients for $a_{i,j}$, $b_i$ and $c_i$ can be found in appendix \ref{ap:rk}.

\subsection{Timestep}
When using an explicit time integrator, the timestep has to fulfill a Courant-Friedrich-Lewy (CFL) condition to achieve numerical stability. The timestep $\Delta t^K$ of the cell $K$ is calculated as \citet{cs5}.

\[
\Delta t^K = \frac{C}{2N_p + 1} \left ( \sum_{i=1}^d \frac{|v_i^K| + c_s^K}{\Delta x_i^K} \right)^{-1},
\]

where $c_s = \sqrt{\gamma p / \rho}$ is the sound speed, $v_i^K$ is the $i^{th}$ component of the velocity average at cell $K$, $\Delta x_i^K$ the mesh-width in the $i^{th}$ dimension.

\subsection{Solution limiters}
It is known that nonlinear equations can develop discontinuities at finite time and that non-physical oscillations develop in the numerical solution in the presence of discontinuities. These, in turn, reduce the pointwise accuracy of the method, lead to loss of convergence near the discontinuity and to the appearance of artificial and persistent oscillations near the discontinuity point \citep{warburton}. Furthermore, for physical systems, it is of interest to have the solution fulfilling certain constraints, such as positivity or boundedness (e.g positive pressure and density). To stabilise the solution, limiters can be used. These, in turn, will affect the quality of the numerical solution.

In this work, we make use of a positivity preserving limiter \citep{ZHANG20108918} to guarantee that the pressure and density remain positive. When using this limiter, there is a further restriction on the timestep, which includes the weight of the first Gauss Lobatto quadrature node, denoted as $w_1$, appropriate for the limiter for a ${N_p}^{th}$ order approximation: 

\[
\Delta t^K = C \min{\bigg( \frac{1}{2N_p + 1},\frac{w_1}{2}\bigg) } \left ( \sum_{i=1}^d \frac{|v_i^K| + c_s^K}{\Delta x_i^K} \right)^{-1}.
\]

\section{Implementation}
\label{sec:implement}
The availability of computational resources such as GPUs has brought renewed interest in compute intensive methods, in which performance is bound by sheer computation rather than by memory access. The discontinuous Galerkin methods having a small stencil and being compute intensive on most platforms fit these requirements. Here we present a two dimensional Cartesian implementation of the discontinuous Galerkin method on GPUs, using CUDA and MPI.

\subsection{Overview of the Algorithm}
The succession of steps of our implementation is as follows:
\begin{enumerate}
    \item{Initial conditions}
    \begin{enumerate}
        \item{Initialize nodes of primitive variables}
        \item{Convert to nodes of conservative variables}
        \item{Integrate to modes of conservative variables}
    \end{enumerate}
    \item{CFL condition: Find global time step }
    \item{Runge-Kuta sub-stepping:}
    \begin{enumerate}
        \item{Compute: volume fluxes, face fluxes, source terms}
        \item{Compute modal update}
        \item{Apply Boundary Conditions}
        \item{Apply Limiters}
    \end{enumerate}
    \item{If $t=t_\text{output}$: Copy modes to CPU and output them}
    \item{If $t<T_\text{end}$: Return to step (ii)}
    \item{End simulation}
\end{enumerate}

The evaluation of the limiting time step is done from the zeroth-order modes, which are the average values for each element. The reduction to obtain the global time step is done over a single block of threads making use of shared memory.
We also make use of the GPU's so-called constant memory to store quadrature values and their respective weights as well as the Legendre polynomials evaluated at these quadrature points.

In order to make an implementation capable of running in parallel over several GPUs there is the need to divide the initial domain, in our case building a sub-domain per GPU. The amalgamation of these sub-domains is then done via \emph{Boundary Conditions}, this allows us to design all the other parts of the code as if each sub-domain were an individual domain with $n_xn_y$ active cells and just one layer of inactive cells or "ghost" cells per side. The information to be communicated consist of the modal values for the conservative quantities. In our implementation (mixing FORTRAN and CUDA) we did not make use of CUDA-Aware MPI instructions to perform device to device memory transfers, therefore we still have room available to increase the performance of our code on multi-GPUs platforms.

From now on we will use the expression ``degree of freedom'' to refer to a single element of resolution, so that the number of degrees of freedom is $n_x\times n_y$ in FARGO3D and $n_x\times n_y\times m^2$ for a DG scheme of order~$m$ (note that the number of values that can be set independently to specify a given configuration is four times larger, since we can specify the surface density, the pressure and the two components of the velocity for each element of resolution).

In our present implementation, all our fields consist of linear arrays, and all our kernels are one-dimensional. The mapping of threads to modes was chosen so as to facilitate memory access without enforcing coalesced transactions: we have one matrix per mode, resulting in $m\times m$ matrices of size $n_x\times n_y$ rather than a unique matrix of size $(m\times nx)\times (m\times n_y)$. Previous experimentation with the automatic data management on the GPU with FARGO3D \citep[see][]{2016ApJS..223...11B} have shown that using pitched memory for multidimensional arrays led to little improvement, if any at all. This is likely due to the two levels of cache available on modern GPUs, as well as sophisticated transaction mechanisms with the global memory on GPUs with recent compute capabilities, leveraging the requirement for alignment, which is no longer so much of a concern since compute capabilities 2.0.

\subsection{GPU performance}
\begin{figure}
\includegraphics[width=\columnwidth]{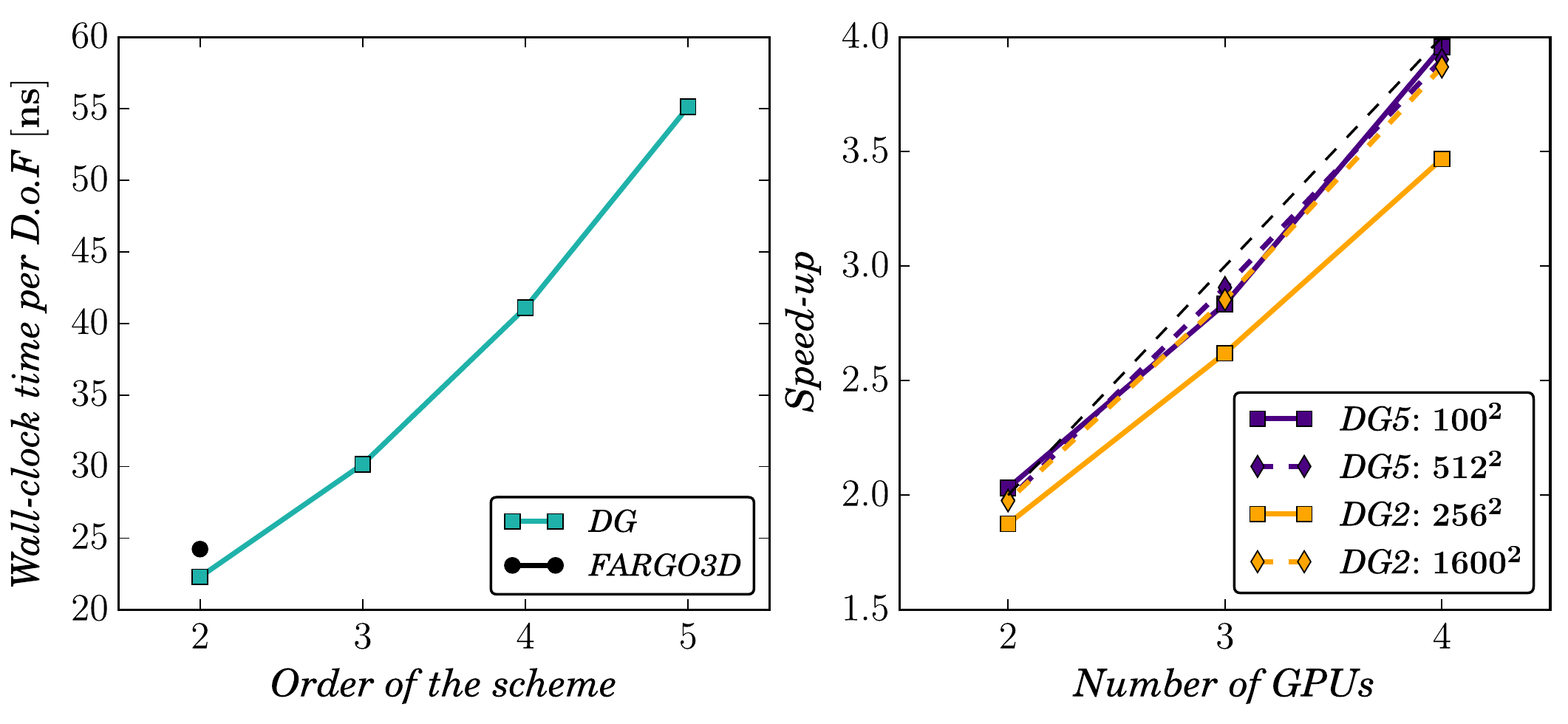}
 \caption{On the left is the average sub-step time taken per degree of freedom as a function of the method's order. On the right is the speed-up observed in the DG code as a function of the number of GPUs used. This data was gathered using NVIDIA's Tesla K20s with error control (ECC) activated.}
 \label{fig:1}
\end{figure}

With the purpose of validating our implementation of the DG method, we measured the wall-clock time of a sub-step with different resolutions and different spatial orders. We then compared these times against the ones measure for FARGO3D. Our results are shown on the left side of figure~\ref{fig:1}, where we plot the average execution time \emph{per degree of freedom} against the order of the scheme. It is important to clarify that this is the wall-clock time for one sub-step, which in DG translates to one of the stages in the \emph{RK} sub-stepping. On the right side we present strong scaling curves to show the performance of the code as a function of the number of GPUs used. We can observe that even though we do not make use of CUDA-Aware MPI instructions we still have a scaling close to the optimal one. We can also see that with higher order it is possible to get closer to optimal scaling even with low resolutions. A weak scaling test showed that we obtain a 60x speed up ratio when running the code on 64 P100 GPUs.

We mention that we have developped over the past year a GPU version only of the code, so we cannot quote an accurate speed up ratio with respect to a CPU core. However, we observed with an earlier, non optimized CPU version of the code, a speed up ratio comprised between 100 and 350 (a larger ratio is obtained at higher order of the scheme, which is likely due to the fact that higher order schemes are most compute intensive). This ratio was obtained respectively with K80 GPUs and Intel\texttrademark{} Xeon E5 cores. 

\section{Test problems}
We present hereafter a number of standard test problems in order to validate our implementation.
\label{sec:benchmarks}
\subsection{Isentropic Vortex}
\label{sec:isentropic}
The isentropic vortex problem describes the convection of an isentropic vortex in an inviscid flow  \citep{yee99}. The physical domain is the square $[0,10]\times[0,10]$, the vortex is centred at $(x_c,y_c) = (5.0,5.0)$,  $r=\sqrt{(x-x_c)^2 + (y-y_c)^2}$ and the boundary conditions are periodic. 
The initial conditions for the primitive variables are:
\begin{align*}
\rho &= \left[ 1 - \frac{(\gamma -1)\beta^2}{8\gamma \pi^2}\exp \big( 1 - r^2\big) \right]^{\frac{1}{\gamma-1}}, \\
v_x &= 1 - \frac{\beta}{2\pi}\exp\left(\frac{1-r^2}{2}\right)(y-y_c), \\
v_y &= 1 + \frac{\beta}{2\pi}\exp\left(\frac{1-r^2}{2}\right)(x-x_c), \\
p &= \rho^\gamma,
\end{align*}
for $\gamma = 1.4$, while the free stream conditions are given by:
$$\rho = 1.0, \quad v_{x,\infty} = 1.0, \quad v_{y,\infty} = 1.0, \quad p = 1.0
$$

\subsubsection{Empirical convergence rate}
The empirical error estimates are calculated using the $\mathcal{L}_1$-error norm:

\begin{equation}
    \mathcal{L}_1 = ||u_h(\bm{x}) - u(\bm{x}) ||_{1}, \quad \bm{x} \in \Omega . 
\end{equation}

It is shown by \citet{shuconvergence} that a convergence rate of $N_p+1$ in $\mathcal{L}_1$-norm is expected for approximate polynomial solutions of degree $N_p$ and smooth enough solutions. This quantity is computed with a suitable numerical quadrature:

\begin{equation}
    ||u_h(\bm{x}) - u(\bm{x}) ||_{1} \approx  \sum_{K \in \Omega}  \frac{1}{4}\sum_{i = 0}^{N_p} \sum_{j = 0}^{N_p}  \mid u_h(x_i,y_j) - u(x_i,y_j) \mid w_i w_j \Delta x \Delta y
 \end{equation}

The system is evolved until $T = 10$ i.e. until the vortex crosses the box and returns to its initial position.

\begin{figure}
 \includegraphics[width=0.9\columnwidth]{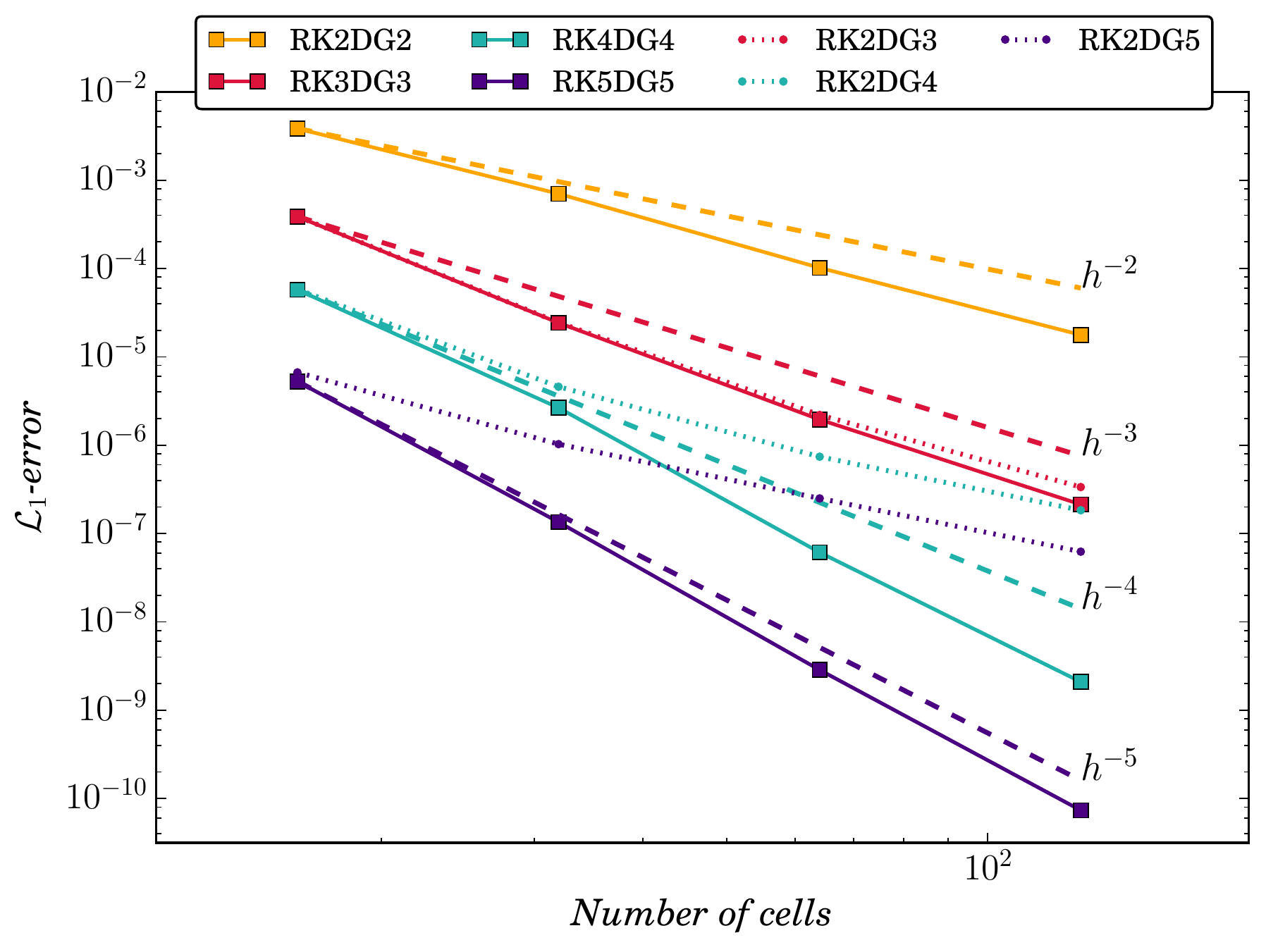}
\caption{Convergence of the RKDG method in $\mathcal{L}_1$-norm for different spatial and time discretization orders for the isentropic vortex case.}
 \label{fig:2}
\end{figure}

As shown in figure~\ref{fig:2}, we observe an empirical convergence rate which is close to the expected theoretical one. We note that reducing the order of the time integration still leads to a decrease in the $\mathcal{L}_1$-norm, although the convergence rate becomes dominated by the time integration error. However, as shown in \citet{truncationerror}, it is possible to recover the right convergence rate if the CFL condition is lowered. In the practical sense, this means that for further experiments, we might be able to reduce the order of the time integration instead of matching the spatial integration order with the time integration order and still attain a low error. This is relevant, as for higher than $4^{th}$ order time integration, it is necessary to have a number larger than the desired order of Runge Kutta sub-steps \citep{rk5}, which becomes prohibitively expensive.

\subsection{Gresho Vortex}
\label{sec:gresho}

The Gresho vortex problem is a rotating steady solution for the inviscid Euler equations \citep{Liska2003}, often used to test conservation of vorticity and angular momentum. The angular velocity $v_\phi$ depends only on the radius and the centrifugal force is balanced by the pressure gradient. The smoothing of the angular velocity profile is a measure of how well the code preserves angular momentum \citep{volker}. The physical domain is defined by $[0,1]\times[0,1]$, the vortex is centred at $(x_c,y_c) = (0.5,0.5)$ and $r=\sqrt{(x-x_c)^2 + (y-y_c)^2}$. The boundary conditions are gradient free:

\begin{equation*}
\nabla u(\vec{x})\cdot \vec{n} \rvert_{\vec{x}\in \partial \Omega} = 0, \quad \mbox{for } u \mbox{ a conserved variable } \rho, v_x, v_y, p.
\end{equation*}

The initial conditions for the primitive variables are:
\begin{equation*}
\rho = 1.0, \quad v_x = -v_\phi \frac{(y-y_c)}{r}, \quad v_y = v_\phi \frac{(x-x_c)}{r}, \quad p = p(r),
\end{equation*}
with the orbital velocity $v_\phi$ and pressure $p$:
\begin{equation*}
    v_{\phi}(r) =  \begin{cases}
    5r & r < 0.2 \\
    2-5r & 0.2\leq r < 0.4 \\
    0 & r\geq 0.4
  \end{cases}
\end{equation*}

\begin{equation*}
    p(r) =  \begin{cases}
    5 + \frac{25}{2}r^2 & r < 0.2 \\
    9 - 4\log(0.2) + \frac{25}{2}r^2 - 20r + 4\log(r) & 0.2\leq r < 0.4 \\
    3 + 4\log(2) & r\geq 0.4
  \end{cases}
\end{equation*}

The angular momentum $\vec{J}$ and vorticity $\vec{\omega} = \nabla \times \vec{v}$ can be written analytically as:
\begin{equation*}
    \vec{J} (r) =  \begin{cases}
    5r^2 & r < 0.2 \\
    2r-5r^2 & 0.2\leq r < 0.4 \\
    0 & r\geq 0.4
  \end{cases} \quad 
    \vec{\omega} (r) =  \begin{cases}
    10 & r < 0.2 \\
    \frac{2}{r} - 10 & 0.2\leq r < 0.4 \\
    0 & r\geq 0.4
  \end{cases}
\end{equation*}

\begin{figure}
\centering
 \includegraphics[width=\columnwidth]{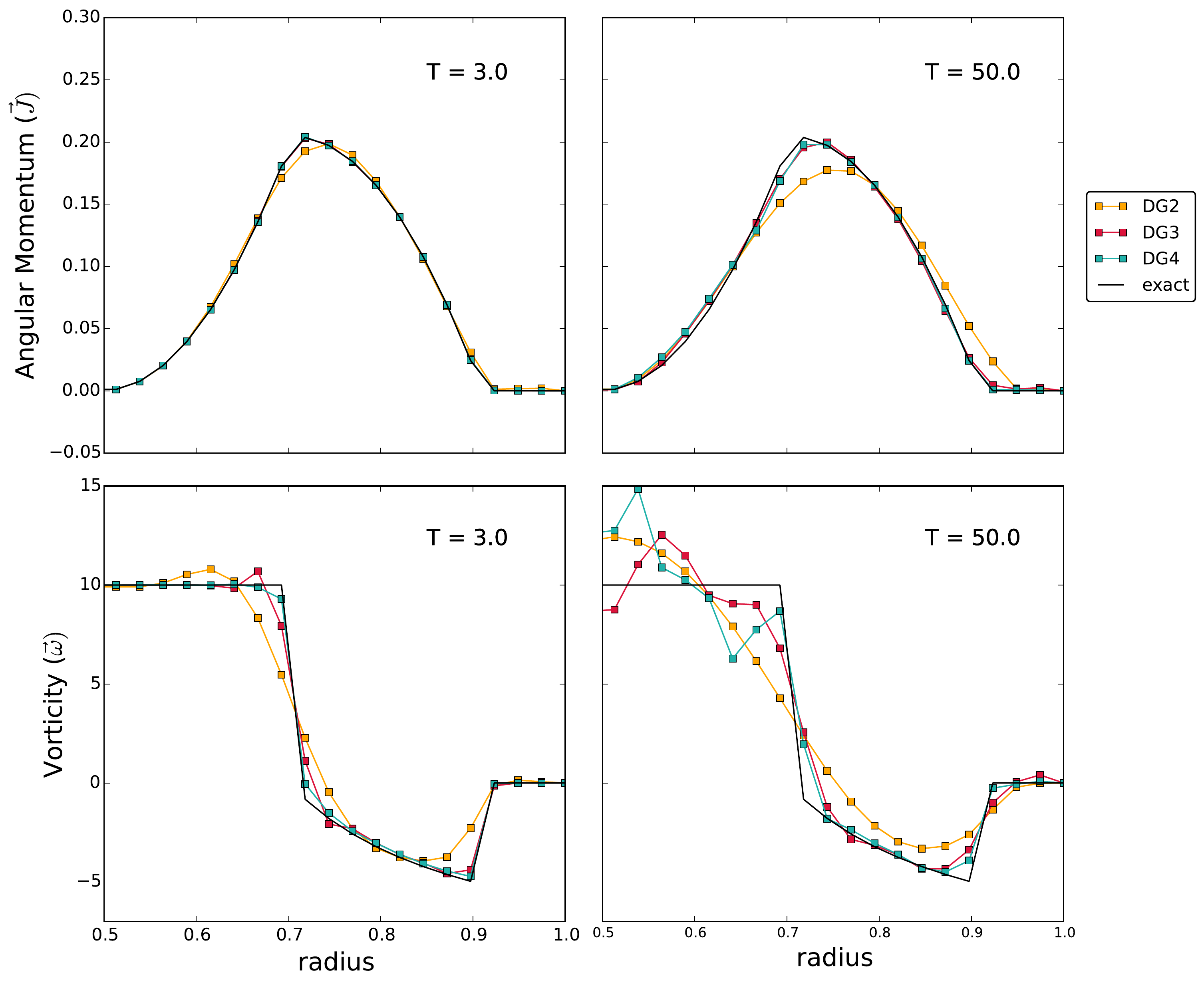}
 \caption{Angular momentum profile at time $T = 3.0$ and $T = 50.0$ for the Gresho's vortex problem for different discretization orders.}
 \label{fig:3}
\end{figure}

It has been shown that the unlimited DG scheme can preserve angular momentum when choosing appropriate basis functions \citep{2015MNRAS.453.4278S}. In figure \ref{fig:3} is shown the profile for the angular momentum at $T = 3.0$ and at $T=50.0$ (corresponding to approximately 2.4 and 40 orbits at $r=0.2$). We note that the angular momentum remains well captured over a longer term evolution, as expected. It has been reported \citep{gresho1,gresho2} that the vortex breaks up for methods which are either too dissipative or unsuitable. However, the vorticity does not behave well over longer term evolution but this is not surprising as the vorticity profile is discontinuous and vorticity is measured is through higher moments of the solution. 

We compare our implementation of the DG scheme with other codes which were benchmarked \citep{Liska2003} with the Gresho vortex case. Following the described setup, we evolve the flow until $T=3$, on a mesh of size $(N_x,N_y)=(40,40)$. 

As shown in table~\ref{table:wendroff}, we find that overall DG methods yield much better results than the other ones for the density (except the second order one, which yields an error comparable to that of PPM or VH1), and an error broadly similar to other methods for the vorticity (except for the fourth order DG scheme, for which the error is typically a factor of two lower than that of PPM or VH1).

\begin{table}
\begin{center}
\begin{tabular}{ | c |c |c |  }
 \hline
 scheme & $\mathcal{L}_1$ vorticity error (\%) & $\mathcal{L}_1$ density error (\%) \\
 \hline
 CFLFh & 20 & 0.16 \\
 JT & 45 & 0.22 \\
 LL & 44 & 0.23 \\
 CLAW & 28 & 0.1 \\
 WAFT & 26 & 0.07 \\
 WENO & 27 & 0.06 \\
 PPM & 13 & 0.04 \\
 VH1 & 15 & 0.04 \\
 \textbf{DG2} & \textbf{20.50} & \textbf{0.05} \\
 \textbf{DG3} & \textbf{12.00} & \textbf{0.01} \\
 \textbf{DG4} & \textbf{6.46} & \textbf{0.008} \\

 \hline
\end{tabular}
\end{center}
\caption{\label{table:wendroff} Relative $\mathcal{L}_1$-norm error for different codes on the Gresho's vortex problem at $T=3$.}
\end{table}

\section{Protoplanetary disc with an embedded planet}
\label{sec:ppdisc}
\subsection{Setup}
We devised a simple planet-disc setup in order to perform a comparison between our DG code and FARGO3D. The setup consists of a disc with an internal radius $r_\mathrm{in}=0.4$ and an external radius $r_\mathrm{ex}=1.75$, an initially uniform surface density  $\Sigma_0=1$ and an initially uniform pressure $p_0=2.5\times 10^{-3}$. A planet is on a fixed circular orbit at $r_p=1$. Since there is no gradient of pressure and density at the planet's orbit, there is neither a gradient of entropy nor of temperature, and we therefore expect that the corotation torque acting on the planet is only the vortensity related corotation torque \citep[eg][and refs. therein]{2017MNRAS.471.4917J}. The adiabatic sound speed at the planet location is therefore $c_s^\mathrm{adi}=\sqrt{\gamma p_0/\Sigma_0}\approx 0.059r_p\Omega_p$ (where $\Omega_p$ is the planet's orbital frequency), while the isothermal sound speed is $c_s^\mathrm{iso}=0.05r_p\Omega_p$, which corresponds to a pressure scale-length $H=0.05r_p$, hence the disc's aspect ratio at the planet location is $h=H/r_p=0.05$. The planet mass is $M_p=6.0\times 10^{-5}M_*$, where $M_*$ is the mass of the central object\footnote{This would translate into a $20\;M_\oplus$ planet for a central mass equal to that of the Sun.}. The planet's gravitational potential has a smoothing length $\epsilon_p=0.03r_p$. 
Since our frames, both in our DG codes and in FARGO3D, are centred on the star, they are not strictly inertial, the star being accelerated by the planet and the disc. This gives rise to an additional term in the gravitational potential, called the indirect term. This term is in general minute and it is not crucial for the comparison that we undertake, so we discard it hereafter in the two codes.
We use the unit system in which $M_*$ is the mass unit, $r_p$ the length unit and $\Omega_p^{-1}$ the time unit, which implies that in this unit system the gravitational constant $G$ is unitary\footnote{Should we take into account the indirect term, we should rather take $M_*+M_p$ as the mass unit for this statement to hold.}.
For the \textit{DG} setup we use a square box with a side of length $4.5r_p$ (going from $-2.25r_p$ to $2.25r_p$).

The fields are initialized as
\begin{align*}
\Sigma &= \frac{\Sigma_0}{1+f(r)}\\
v_x &= -v_\phi \frac{y}{r} \\
v_y &= v_\phi \frac{x}{r} \\
p &= \frac{p_0}{1+f(r)},
\end{align*}
where we chose $f(r)=\exp\left(\frac{r-r_\mathrm{ex}}{r_ph}\right)$ to provide a smooth transition at the outskirts of the disc. The angular velocity $v_\phi$ that leads to rotational equilibrium with this density profile is
\begin{align*}
	v_{\phi}^2 &= r^2\Omega^2 +\frac{r\partial_r({c^2_\mathrm{s,iso}}\Sigma)}{\rho}=
    \Omega^2 -\frac{rc^2_\mathrm{s,iso} f(r)}{r_ph[1+f(r)]},
\end{align*}
where the ratio $\gamma$ of specific heats is set to $1.4$. For the orbital frequency $\Omega$ we have solid rotation inside an inner limit and a Keplerian flow elsewhere: 
\begin{equation*}
    \Omega^2 (r) =  \begin{cases}
    \frac{GM_*}{r_\mathrm{in}^3} & r \leq r_\mathrm{in} \\
    \frac{GM_*}{r^3} & r > r_\mathrm{in}.
    \end{cases}
\end{equation*}

The setup also includes wave-killing boundary conditions as described in \cite{2006MNRAS.370..529D}. A field $Q$ is dampened towards its unperturbed value $Q_0$ every time step according to the following prescription:
\begin{equation*}
	Q=\frac{\Delta t Q_0 + \tau Q}{\tau+\Delta t},
\end{equation*}
the damping time being
\begin{equation*}
\tau = 2\pi \sqrt{\frac{r_d^3}{GM_*}}\times\frac{1}{R(r)},
\end{equation*}
where the ramp function 
 \begin{equation*}
R(r)=\left(\frac{r-r_d}{r_\mathrm{in/ex}-r_d}\right)^2
\end{equation*}
is chosen to span the interval $0$ to $1$ with a parabolic behaviour in the zone from $r_d$ to the boundary radius $r_\mathrm{in/ex}$. The damping radius $r_d$ is chosen for the internal boundary as
\begin{align*}
	r_d = r_\mathrm{in}1.15^{3/2}
\end{align*}
and for the external one as
\begin{align*}
r_d &= r_\mathrm{ex}1.15^{-3/2}.
\end{align*}
The dampened fields are the density, velocities and temperature.

\subsection{Results comparison}
\begin{figure*}
 \includegraphics[width=.95\textwidth]{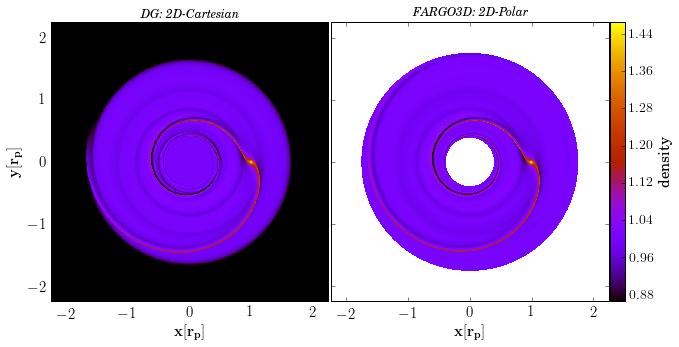}
 \caption{Snapshots of the surface density for an adiabatic disc $50$ orbits after the insertion of the planet. On the left we plot results of the RK2DG5 scheme with a $640\times 640$ Cartesian mesh whereas on the right we plot results of FARGO3D with a $3574\times 768$ polar mesh and orbital advection.}
 \label{fig:4}
\end{figure*}

We present the results of the Cartesian \emph{DG} code and those of \emph{FARGO3D} with a polar grid, both codes performing numerical simulations of the planet-disc setup.  We undertook simulations with increasing resolution and order for the \emph{DG} code. 
In Fig.~\ref{fig:4} we show a comparison of the surface density map obtained with a RK2DG5 scheme and another obtained with FARGO3D. The location and contrast of the spiral wake is almost undistinguishable between the two runs as long as one stands away from the boundaries. Besides, the region where the torque originates is located relatively far from the region that limits the timestep through the CFL condition. As a consequence, the effective Courant number of this specific region is small, which results in minute differences at a given location from one timestep to the next, and thus different time order schemes yield very similar results.

For the whole set of runs we monitored the specific torque exerted onto the planet position $\vec{r}_p$:
\begin{equation}
\vec{\Gamma}=\sum_{n}\vec{r}_p \times \vec{g}_{n}=\vec{r}_p\times \sum_{n}\frac{GM_{n}(\vec{r}_{n}-\vec{r}_p)}{\left[(\vec{r}_{n}-\vec{r}_p)^2+\epsilon^2_p\right]^{3/2}},
\end{equation}
where $M_{n}$ and $\vec{r}_{n}$ represent respectively the mass and position  of cell $n$, whereas $\vec{g}_n$ represents the acceleration imparted by the material of cell $n$ at the planet's location.  The evaluation of this acceleration includes the planet's smoothing length $\epsilon_p$. Given that we are considering a two-dimensional case, the only non-zero component of $\vec{\Gamma}$ is the $z$-component. From here on we will refer to this component as the total torque $\Gamma$.\\

We normalize this torque to $\Gamma_0=\Sigma\Omega^2r_p^4q/h^2$
and from now on quote values of $\gamma\Gamma/\Gamma_0$.\\

\begin{figure*}
 \includegraphics[width=.95\textwidth]{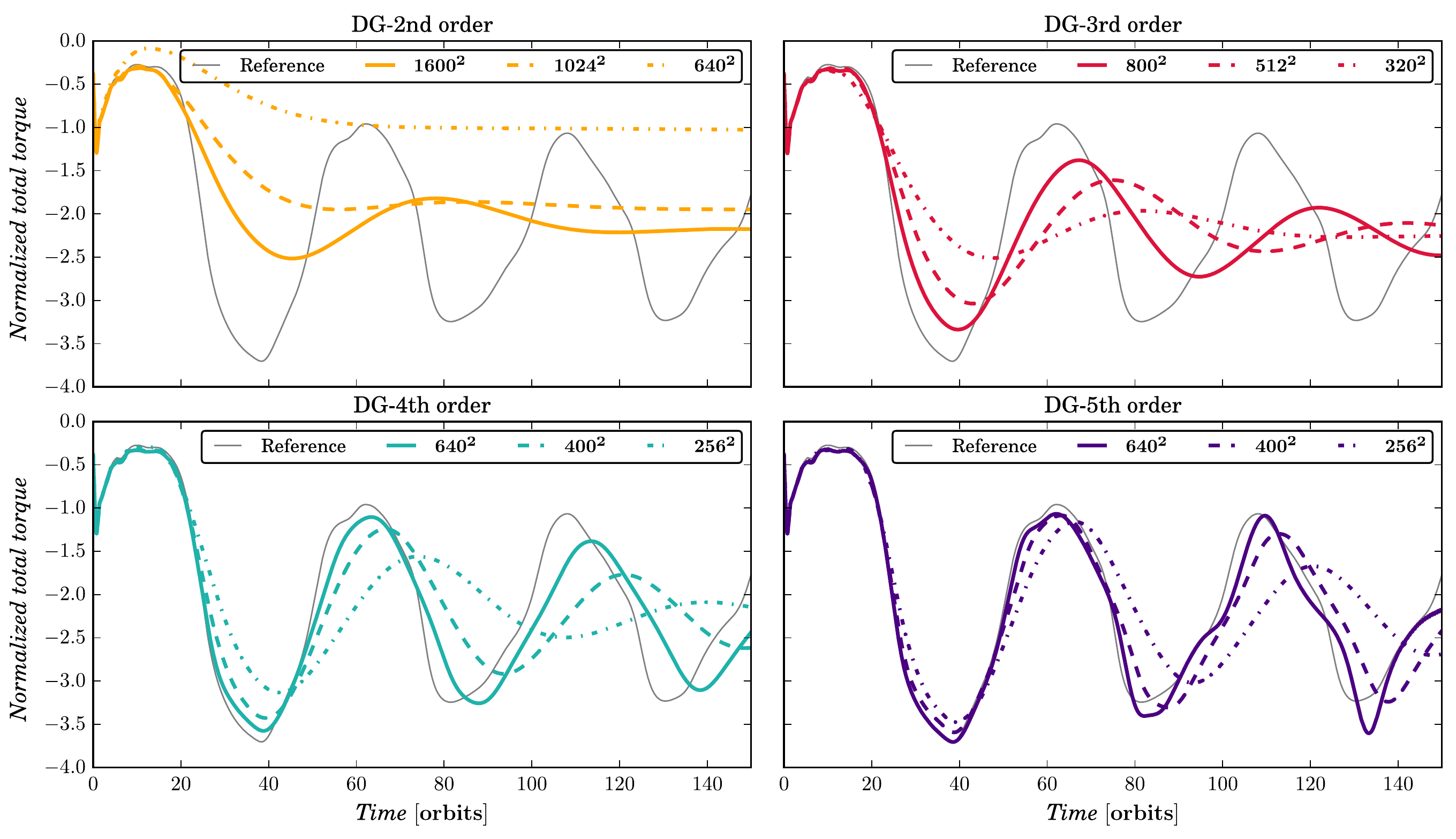}
 \caption{Normalized total torque obtained with the DG code for an adiabatic disc up to $150$ orbits, for different orders of the scheme and different resolutions. Each plot corresponds to a given order, and shows the torque evolution for different resolutions. The reference torque is obtained with FARGO3D using orbital advection. No physical viscosity was included in these calculations, and the departure from the serrated behaviour of the torque is exclusively accounted for by numerical diffusion. }
 \label{fig:5}
\end{figure*}

We start by presenting in figure~\ref{fig:5} the results of the \emph{DG} code, where we show how the response depends on resolution for different orders of the scheme. As we increase the order we  observe a torque exhibiting more and more the serrated behaviour expected for inviscid discs \citep{2007LPI....38.2289W}. \\ 

\begin{figure}
 \includegraphics[width=\columnwidth]{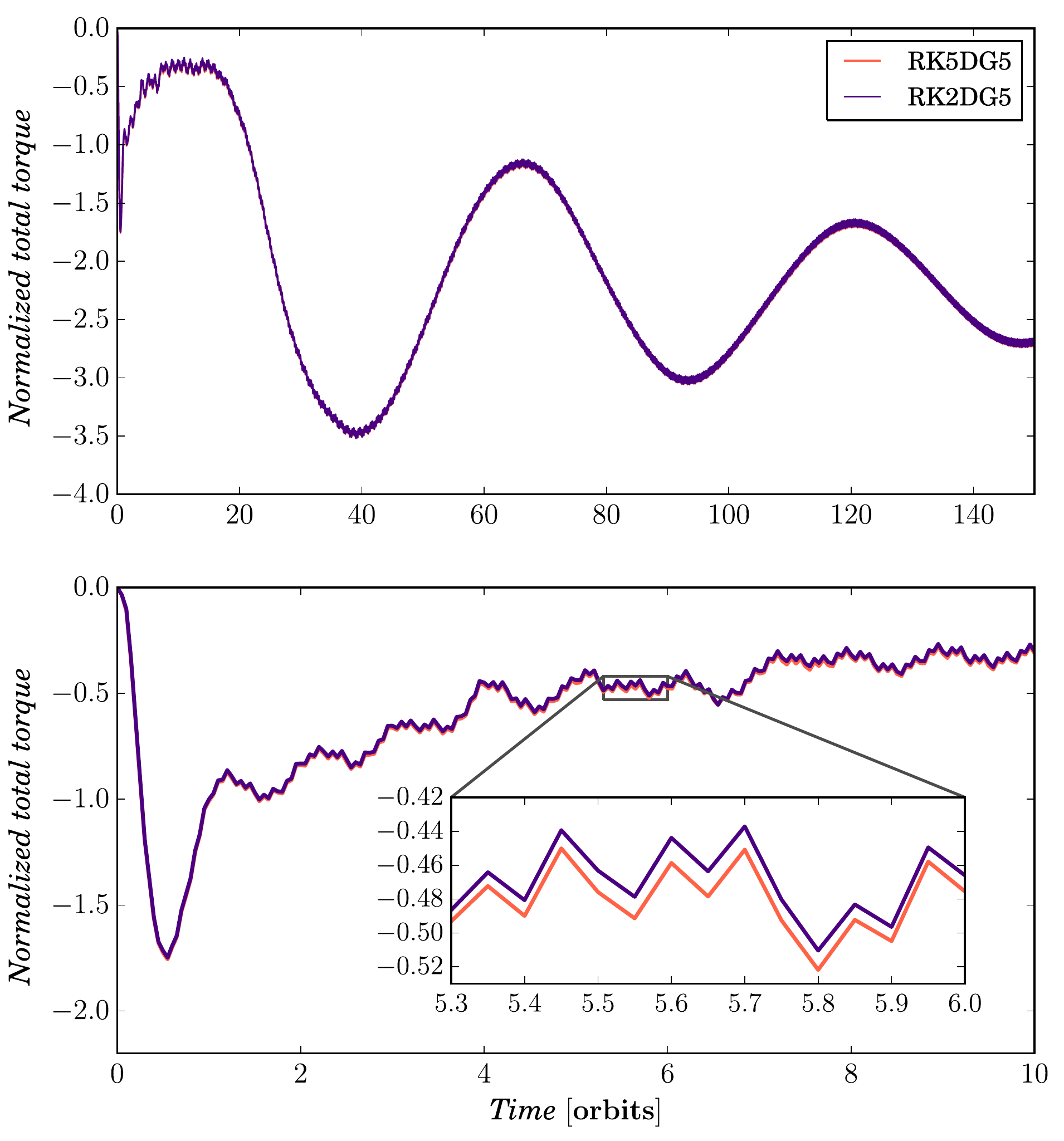}
 \caption{Normalized total torque obtained with the DG code for a $5^{th}$ spatial order with $2^{nd}$ and  $5^{th}$ order time integrators.}
 \label{fig:6}
\end{figure}

In figure~\ref{fig:6} we show the results of the DG5 scheme with RK2 and RK5 time integrators. These results can also be generalized to $3^{rd}$ and $4^{th}$ spatial order schemes, for which we find that a second time order integrator yields virtually undistinguishable torque estimates. This is likely due to the fact that the horseshoe region, from which most of the torque originates, is resolved on a relatively small number of zones. As shown previously in figure~\ref{fig:2}, at low resolution, second order of time integrators led to errors on the norm very similar to higher order time integrators.

\begin{figure}
 \includegraphics[width=\columnwidth]{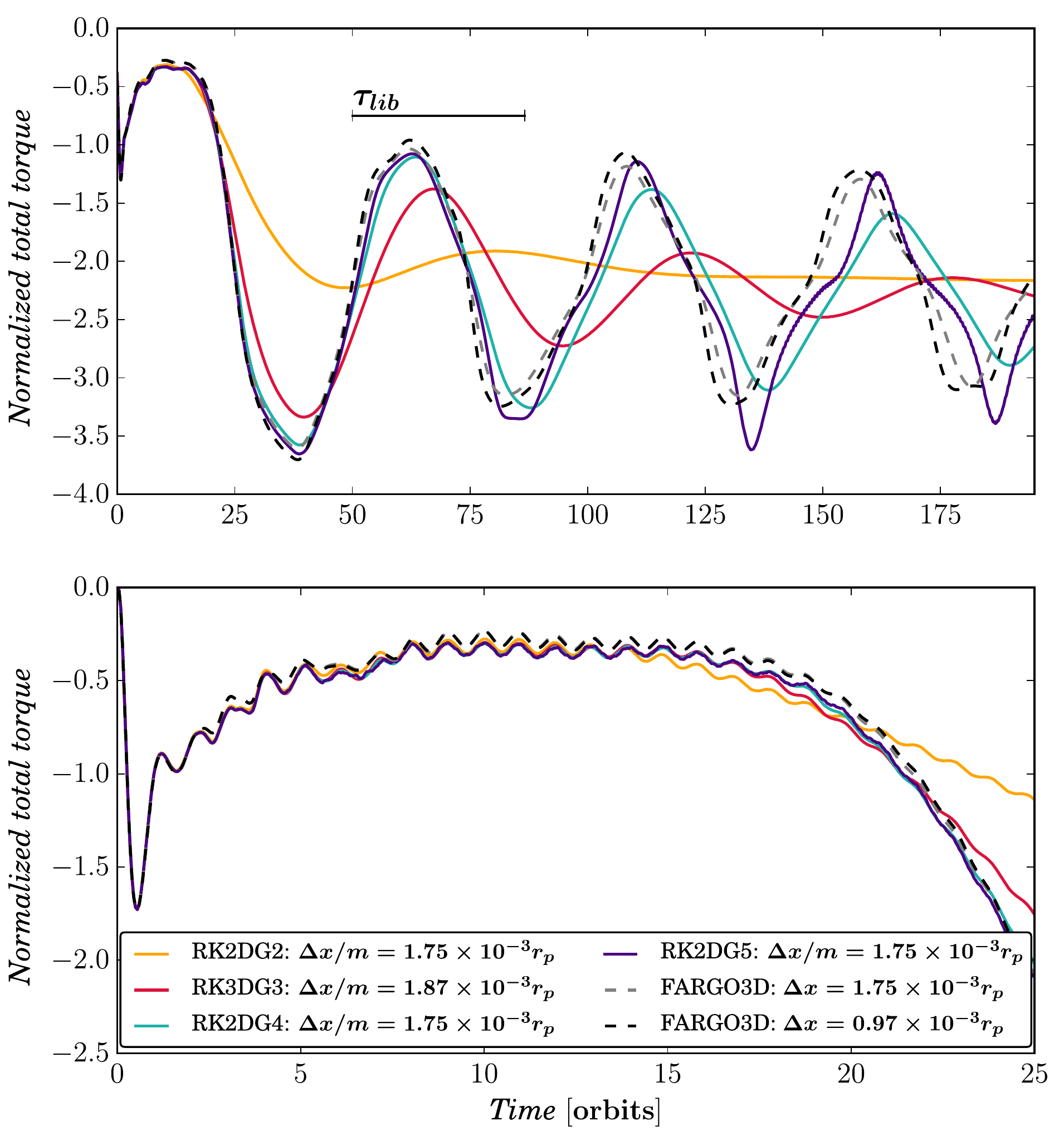}
 \caption{Time dependence of the normalized total torque for an adiabatic disc up to $200$ orbits. In solid lines we plot the results for the \emph{DG} code and with dashed lines the ones for FARGO3D with a polar mesh and including the FARGO scheme. We show two different resolutions of the FARGO3D runs, which are nearly undistinguishable, which shows that FARGO3D results are converged.}
 \label{fig:7}
\end{figure}

Figure~\ref{fig:7} contains the normalized total torque for different orders with similar resolution. Here the resolution for the \emph{DG} code is taken as the cell length over the order of the approximation. The runs for \emph{FARGO3D} were designed to have square cells at $r=r_p$, the first curve here has same cell size as the ones shown for \emph{DG} with second, fourth and fifth order.\\

We observe a  behaviour  resembling more closely analytical expectations for higher orders with an equal number of degrees of freedom. The torques in the top plot of figure~\ref{fig:7} show a high degree of similitude between both codes, especially for the highest orders of the DG scheme. The similitude even holds if we focus on the high frequency, low amplitude components of the torque at early stages  (bottom plot of figure~\ref{fig:7}) where the oscillations in both codes have a similar structure.
Given the marked difference between the two codes (both in numerical method and mesh geometry), this strongly suggests that this behaviour is of physical origin rather than being a numerical artefact.

\subsection{Estimated numerical viscosity}
\begin{figure}
 \includegraphics[width=\columnwidth]{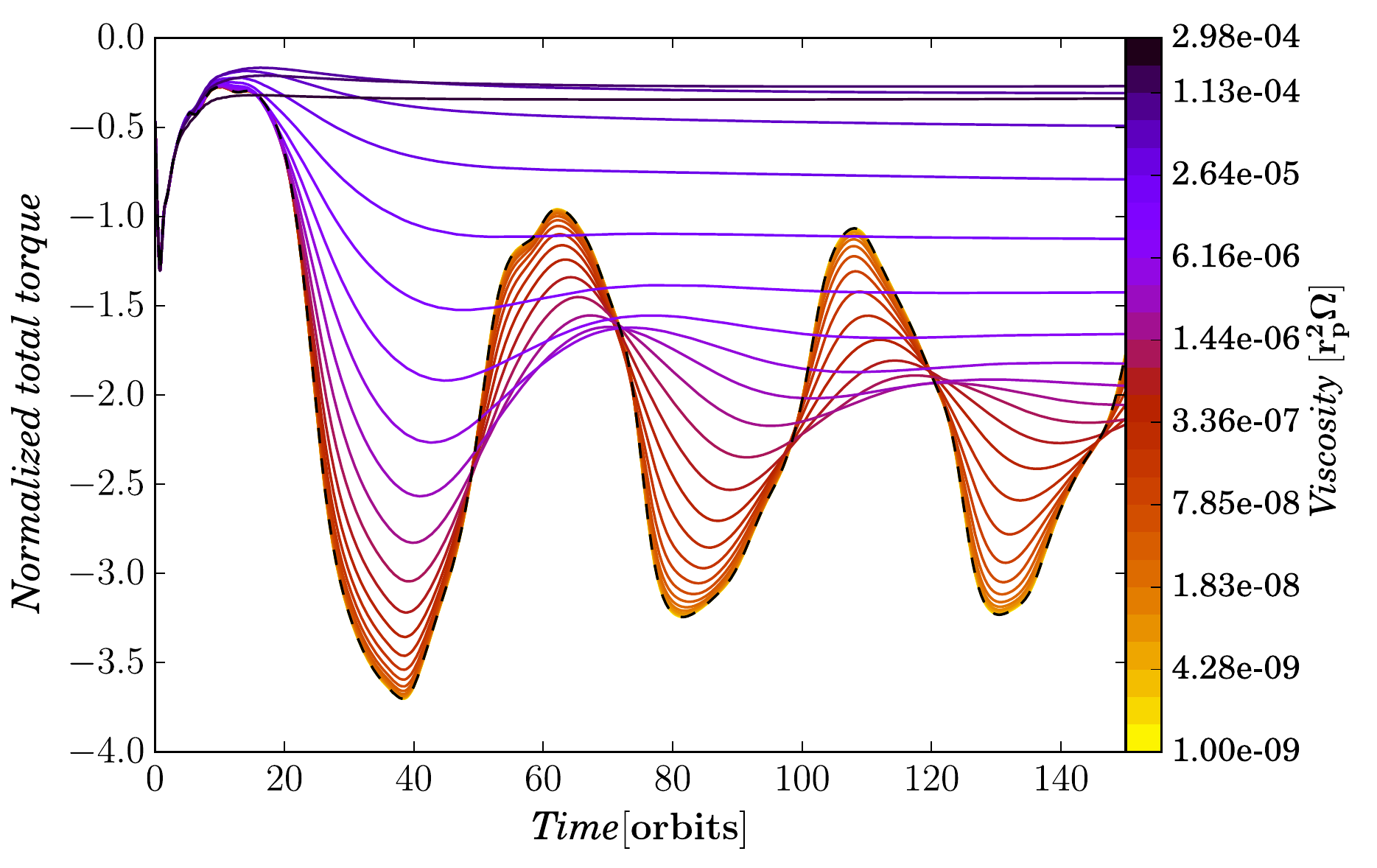}
 \caption{Time dependence of the normalized total torque for adiabatic discs with increasing kinematic viscosity $\nu$. These results are obtained with FARGO3D with a mesh of dimensions $6444\times 1384$ respectively for azimuth and radius. The runs were performed in the co-rotating frame with orbital advection.}
 \label{fig:8}
\end{figure}

Horseshoe dynamics, which give rise to the corotation torque, can be essentially reduced to an advection and diffusion problem, in which the advection stems from the Keplerian flow and the diffusion comes here from the numerical scheme itself.
In order to quantify the numerical viscosity of the DG method we resort to a comparison with high resolution simulations performed with the FARGO3D code using orbital advection. Since FARGO3D solves the Navier-Stokes equations, we run a set of simulations spanning kinematic viscosities $\nu$ from $10^{-9}r_p^2\Omega_p$ to $10^{-4}r_p^2\Omega_p$. We then obtain the palette of torques presented in figure~\ref{fig:8}, to which we can compare the results of our DG code in order to assess its numerical viscosity for a given resolution and scheme order.

\begin{figure}
	\includegraphics[width=\columnwidth]{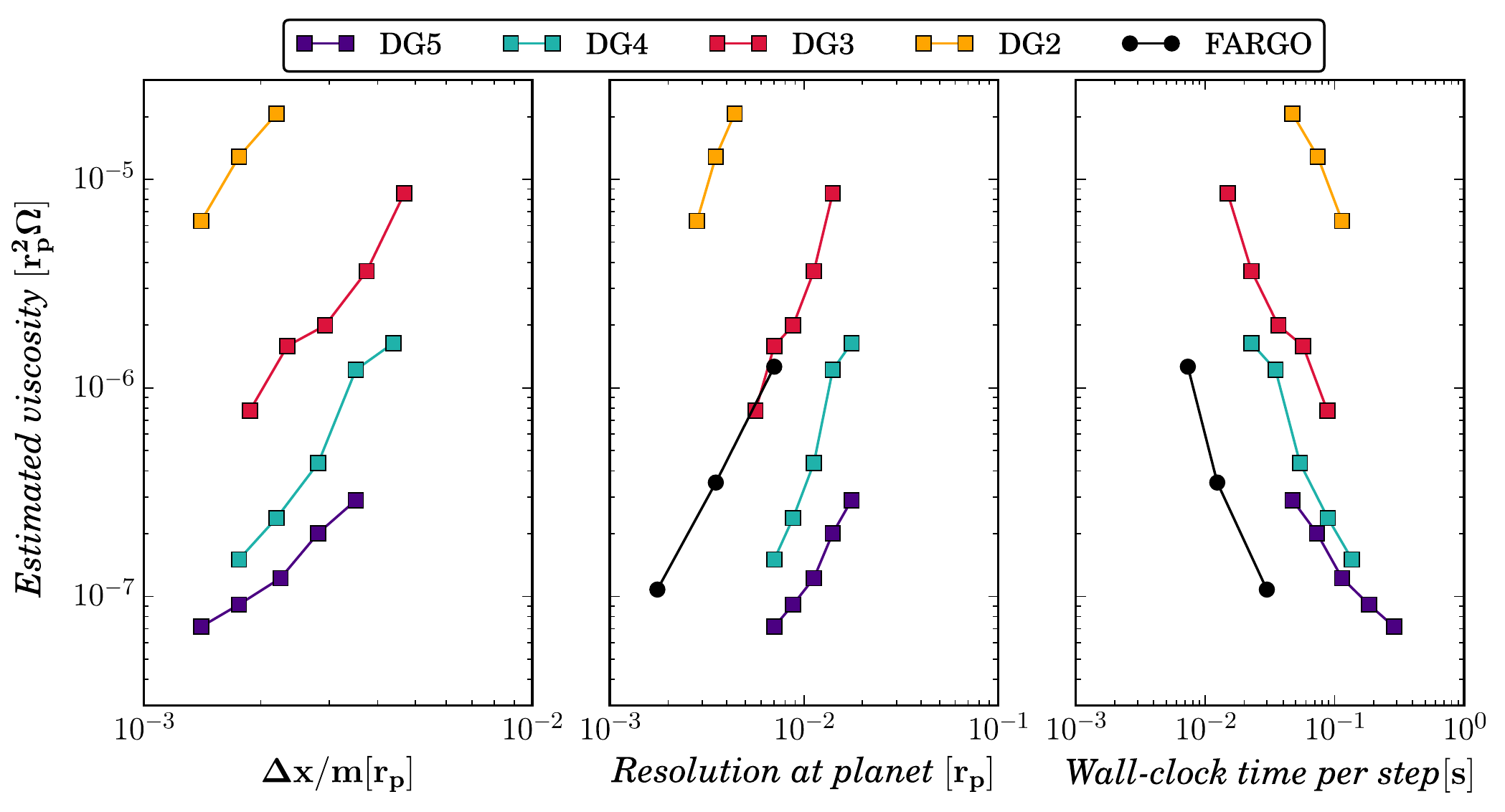}
    \caption{Numerical viscosity of the \emph{DG} schemes inferred from the viscosity palette obtained with FARGO3D.The results are for a second order Runge-Kutta time integrator for \emph{DG}, and orbital advection and a non-rotating frame for FARGO3D. The left plot shows the inferred viscosity for \emph{DG} as a function of the effective resolution $\Delta x/m$. The centre plot shows the viscosity as a function of the resolution $\Delta x$ at the planet position. The right plot shows the viscosity as a function of the wall-clock time per step, which shows that FARGO3D is more efficient than DG schemes, at least up to order~5.}
    \label{fig:9}
\end{figure}

\cite{2010ApJ...723.1393M} find that successive maxima and minima of the corotation torque for low values of the shear viscosity are approximately in geometric sequence. We extract the  ratio $\tilde{\rho}$ of this sequence using the first three extrema of our low viscosity runs (up to $\nu \approx 10^{-6}r_p^2\Omega_p$). This value of $\tilde{\rho}$ is what we use to match the physical viscosity of FARGO3D runs to the numerical viscosity of DG runs. Namely, from the values if $\tilde{\rho}$ obtained from the FARGO3D runs we build a polynomial approximation of the relation between $\tilde{\rho}$ and the viscosity $\nu$. This relation, in turn, is used backwards to get a viscosity estimate for a given value of $\tilde{\rho}$ measured in a DG run. The estimates of the numerical viscosity are shown in figure~\ref{fig:9}. We stress that our method yields an accurate evaluation of the numerical viscosity of a given scheme in the very specific problem of horseshoe dynamics. Should an observable other than the torque be used to infer the numerical viscosity of a scheme (such as the minimum density in a gap opening situation, for instance), a different value could be found.  Owing to the exquisite sensitivity of the ultimate torque value on the effective viscosity, we believe that our method provides a fair estimate of the scheme's intrinsic viscosity which can be in turn used to assess the scheme's properties in widely different situations, in particular those for which the dominant effect of the scheme properties is a diffusion of vortensity.

The left plot of figure~\ref{fig:9} shows the advantage, in term of numerical viscosity, of increasing the scheme's order, for a given effective resolution (size of a cell divided by the scheme order). The centre plot shows that FARGO3D with orbital advection performs nearly as \emph{DG} with $3^{rd}$ order with respect to the resolution at the planet position, while the $4^{th}$ and $5^{th}$ order \emph{DG} schemes outperform FARGO3D, despite the considerably less favourable mesh geometry and the lack of orbital advection. With respect to execution time, the right plot shows that FARGO3D outperforms \emph{DG}, by nearly an order of magnitude.

\section{Conclusions}
\label{sec:conclusions}
We have shown the applicability of the discontinuous Galerkin methods to simulations of planet-disc interactions, being able to obtain negligible numerical viscosities with high-order schemes. We have shown that for a given number of degrees of freedom we reach lower viscosities by increasing the order of the scheme rather than by increasing the resolution. The \emph{DG} code with a Cartesian mesh and a non-rotating frame is able to reproduce the results of FARGO3D, for which we need a polar mesh and either orbital advection or a frame co-rotating with the planet to properly capture the disc's torque. We note that the effective viscosities of protoplanetary discs may be extremely small. Many observations suggest the existence of vortices, the persistence of which requires a parameter of Shakura-Syunyaev $\alpha$ of at most $10^{-4}$ \citep{2016MNRAS.458.3918Z}, which translates in our setup into $\nu=2.5\cdot 10^{-7}r_p\Omega_p^2$. Besides, it has been suggested that angular momentum transport driving accretion in protoplanetary discs might not be of viscous nature \citep{2017ApJ...837..163R}, which stresses the need for numerical methods with very low numerical viscosity. Our \emph{DG} code is slower than FARGO3D and therefore is probably of little use for single disc setups. Also, being at the present time two-dimensional, it should essentially be regarded as a proof of concept. It nevertheless strongly suggests that DG schemes may be very useful in more complex situations when low-viscosity flows must be captured, such as multi-scale simulations of protoplanetary discs and their environment, for which Cartesian AMR are a tool of choice. 




\bibliographystyle{mnras}



\appendix

\begin{table*}
\centering
{
$\begin{array}{l|lllll}
0\\
0.39175222700392 & 0.39175222700392 & & & & \\
0.58607968896779 & 0.21766909633821  & 0.36841059262959 & & & \\
0.47454236302687 & 0.08269208670950  & 0.13995850206999 &  0.25189177424738 &  & \\
0.93501063100924 & 0.06796628370320  & 0.11503469844438 &  0.20703489864929 & 0.54497475021237 & \\
\hline
& 0.14681187618661 & 0.24848290924556 & 0.10425883036650 & 0.27443890091960 &  0.22600748319395\\
\end{array}$
}
\caption{\label{butcher} Runga-Kutta Butcher tableau for the SSP-RK scheme RK(4,5).\label{tab:ssp45}}
\end{table*}

\begin{table*}
\centering
{
$\begin{array}{l|llllll}
0\\
1/5 & 1/5 & & & & \\
3/10 & 3/40  & 9/40 & & & \\
3/5 & 3/10 & -9/10 & 6/5 &  & \\
1 & -11/54  & 5/2 &  -70/27 & 35/27 & \\
7/8 & 1631/55296  & 175/512 &  575/13824 & 44275/110592 & 253/4096 \\
\hline
& 37/378 & 0 & 250/621 & 125/594 & 0 & 512/1771 \\
\end{array}$
}
\caption{6-stage Cash-Karp Butcher tableau for $5^{th}$ order accuracy \citep{cashkarp}.\label{tab:65ts}}
\end{table*}

\section{Timestepping coefficients}
\label{ap:rk}

To perform the time integration a Runge-Kutta method is used. For the ODE:

\[ \frac{d}{dt} u_h = \mathcal{L}(u), \]

and a suitable initial condition $u_h^0$, we obtain the solution at $t^{n+1}$:

\begin{align*}
u_h^{n+1}=u_h^{n}+h\sum_{i=1}^k b_i k_i
\end{align*}
where 
\begin{align*}
k_i &= \mathcal{L}(t^n+c_i\cdot h,y_i+h(a_{i,1}k_1 + ... + a_{i,i-1}k_{i-1}))
\end{align*}

To specify a particular timestepping method, one needs to specify the number of stages $k$ and the coefficients $a_{i,j}$, $b_i$ and $c_i$. This section contains the Butcher tableaus for the different Runge-Kutta timestepping algorithms. The generic Butcher tableau can be seen in table \ref{tab:generic}, and shown in tables \ref{tab:ssp22}, \ref{tab:ssp33}, \ref{tab:ssp45} and \ref{tab:65ts} we have the second, third, fourth and fifth order time integration algorithms, respectively.

\begin{table}
\centering
$\begin{array}{l|lllll}
0\\
c_2 & a_{2,1} & & & & \\
c_3 & a_{3,1}  & a_{3,2} & & & \\
... & ...  & ... &  ... &  & \\
c_k & a_{k,1}  & a_{k,2} &  ... & a_{k,k-1}& \\
\hline
& b_1 & b_2 & ... & b_{k-1} &  b_k\\
\end{array}$
\caption{Generic Butcher tableau for k-stage explicit Runge Kutta method.\label{tab:generic}}
\end{table}

\begin{table}
\centering
$
\begin{array}{l|ll}
0\\
1/2 & 1/2\\
\hline
& 1/2 & 1/2 \\
\end{array}
$
\caption{\label{butcher} Runga-Kutta Butcher tableaus for the SSP(2,2) scheme.\label{tab:ssp22}}
\end{table}

\begin{table}
\centering 
$\begin{array}{l|lll}
0\\
1 & 1 & & \\
3/4& 1/4 & 1/4\\
\hline
&1/6 & 1/6 & 2/3\\
\end{array}$
\caption{\label{butcher} Runga-Kutta Butcher tableau for the SSP(3,3) schemes.\label{tab:ssp33}}
\end{table}


\bsp	
\label{lastpage}
\end{document}